# On the Critical Behavior of the Uniform Susceptibility of a Fermi Liquid Near an Antiferromagnetic Transition with Dynamic Exponent $z = 2$

L. B. Ioffe

*Department of Physics Rutgers University Piscataway, NJ 08855*

A. J. Millis

*AT&T Bell Laboratories 600 Mountain Avenue Murray Hill, NJ 07974*

We compute the leading behavior of the uniform magnetic susceptibility, $\chi$, of a Fermi liquid near an antiferromagnetic transition with dynamic exponent $z = 2$. Our calculation clarifies the role of triangular "anomaly" graphs in the theory and justifies the effective action used in previous work[1]. We find that at the $z = 2$ critical point of a two dimensional material, $lim_{q \to 0} \chi(q, 0) = \chi_0 - DT$ with $\chi_0$ and $D$ nonuniversal constants. For reasonable band structures we find that in a weak coupling approximation $D$ is small and positive. Our result suggests that the behavior observed in the quantum critical regime of underdoped high-$T_c$ superconductors are difficult to explain in a $z = 2$ theory.

The anomalous magnetic behavior of the underdoped high $T_c$ materials has kindled interest in the properties of antiferromagnetic quantum critical points. The data are most clear in $YBa_2Cu_4O_8$. In this compound the planar copper nuclear relaxation rate $1/^{Cu}T_1T \propto 1/T^2$ for $200K \leq T \leq 600K$ while the uniform static susceptibility $\chi(T) = \chi_0 + DT^3$ in the same temperature range. The $T$-dependence of the copper $1/T_1T$ in high $T_c$ materials has been interpreted[4] as due to strong, temperature dependent antiferromagnetic correlations, as expected if the spin dynamics is controlled by proximity to a $T = 0$ antiferromagnetic critical point. Two possibilities have been proposed for the critical point; these are distinguished by the value of the dynamic exponent, $z$. One class of models is based on the $z = 2$ critical point expected for an antiferromagnetic instability of a fermi liquid system[5]. The value $z = 2$ arises because in a Fermi liquid the dominant contribution to the dynamics of an antiferromagnetic spin fluctuation is decay into a particle-hole pair. The $z = 2$ theory is argued[4,6] to apply to the high $T_c$ materials because they are electrical conductors and presumably have fermi surfaces and therefore particle-hole continua into which spin fluctuations may decay. An alternative point of view[7,8] is that the correct model involves undamped (or weakly damped) spin waves with dispersion relation $\omega = ck$, with the spin-wave velocity $c$ remaining constant near the critical point, implying a dynamical exponent $z = 1$. For this point of view to be correct it is necessary either that the particle-hole continuum does not exist or that it does not couple to the spin waves.

One phenomenological argument for the applicability of the $z = 1$ theory to $YBa_2Cu_4O_8$ involves the temperature dependence of the uniform susceptibility, $\chi$. Indeed, the result $\chi(T) \sim AT$ has been demonstrated theoretically for the quantum critical regime of the $z = 1$ model[8], and the coefficient $A$ was shown to be universal (after the spin wave velocity and the $g$-factor are fixed). The rough (factor-of-two) agreement between the predicted value of $A$ and the experimental results for $d\chi/dT$[9] has led several authors[7,8] to propose that the experimental susceptibility should be viewed as the sum of two terms: a weakly temperature dependent background term coming from a fermi liquid like continuum and a term proportional to $T$ coming from quantum critical $z = 1$ dynamics. This point of view has, however, been questioned on phenomenological grounds[6].

In this communication we give a calculation of the critical behavior of $\chi$ near a $z = 2$ antiferromagnetic critical point. In addition we clarify two related theoretical issues: the role of triangular "anomaly" graphs in the theory of the $z = 2$ transition and the role of "gauge invariance" arguments relating the response of a system in a uniform magnetic field to a stiffness against changes in boundary condition in the time direction[10].

Our theoretical approach follows previous work[1,5]: we integrate out the conduction electron degrees of freedom at fixed values of the staggered magnetization $\vec{\phi}$ and externally applied field $\vec{h}_{ext}$, obtaining an expression for the effective action $S^{eff}$, as a functional of $\vec{\phi}$ and $\vec{h}_{ext}$. We then evaluate the free energy implied by $S^{eff}$ using standard diagrammatic and renormalization group means. The result obtained in previous work at $\vec{h}_{ext} = 0$ is

$$S^{eff} = \frac{1}{2} \sum_{k,\Omega} \chi^{-1}(k,\Omega) \vec{\phi}(Q+k,\Omega) \cdot \vec{\phi}(-Q-k,-\Omega]$$
$$+ U \sum_{k_1..k_4, \Omega_1...\Omega_4} \delta(k_1+k_2+k_3+k_4)\delta(\Omega_1+\Omega_2+\Omega_3+\Omega_4) \qquad (1)$$
$$\times [\vec{\phi}(Q+k_1,\Omega_1) \cdot \vec{\phi}(-Q+k_2,\Omega_2)][\vec{\phi}(Q+k_3,\Omega_3) \cdot \vec{\phi}(-Q+k_4,\Omega_4)] + ...$$

where $\chi^{-1}$ is the inverse of the staggered susceptibility and is given by

$$\chi^{-1}(k,\Omega) = \frac{|\Omega|}{\Gamma} + k^2 + \delta \qquad (2)$$

Here $\Gamma$ is a relaxation time, $\delta$ measures the distance to the antiferromagnetic critical point, wavevectors are measured with respect to the ordering wavevector $\vec{Q}$, and both the $\phi^4$ coupling $U$ and higher couplings are irrelevant (or marginally irrelevant) under renormalization in $d \geq 2$ spatial dimensions.

The diagrams contributing to $S_{eff}$ at $h_{ext} = 0$ are shown in Figs. 1a and 1b. Diagrams with an odd number of $\phi$ vertices are forbidden by momentum conservation if the momenta of the $\phi$ fields are restricted to be near the ordering wavevector $Q$. The diagram in Fig. 1a is the sum of all diagrams which are irreducible with respect to the fermion-fermion interaction; this differs from the inverse susceptibility by an additive constant.

The $\phi^4$ coefficient $U$ is given by Fig. 1b and is an analytic function of its arguments if the ordering wavevector $Q$ is not an "extremal vector" of the Fermi surface. By "extremal vector" we mean one which connects two Fermi surface points with parallel tangents. For a circular Fermi surface, any $Q = 2p_F$ would be an extremal vector. The case where $Q$ is an extremal vector will be discussed elsewhere[11].

We now derive $S_{eff}$ in a nearly uniform external applied field. Before presenting the explicit calculations we recall arguments based on rotational invariance, which provide a useful check on our results. The essential point is that spins precess in an applied field, so the equation of motion of a spin $S_i$ in a field $h$ is

$$\partial \vec{S}_i/\partial t = \vec{h}(R_i,t) \times \vec{S}_i + i[H,S] \tag{3}$$

where $H$ is the Hamiltonian in zero field. If the applied field is spatially uniform (but with an arbitrary time dependence) the all spins precess at the same rate, so if $H$ is rotationally invariant the explicit dependence on field in the equations of motion may be removed by the rotation $S_i^a(t) = R_h^{ab}(t)S_i^b(t)$ with

$$R_h^{ab}(t) = \left[T\,exp \int_0^{t'} dt\, \epsilon^{abc}h^c(R_i,t')\right] \tag{4}$$

Therefore the time evolution operator in the presence of the field $h$, $U_h(t)$, is given by (note $\mathbf{R}_h(0) = 1$)

$$U_h(t) = \mathbf{R}_h(t)U_{h=0}(t) \tag{5}$$

where $U_{h=0}(t) = e^{iHt}$. ¿From Eq. (5) it follows that the dependence of the time evolution on a spatially uniform field is only via boundary conditions. This line of reasoning applies also to the imaginary time formalism used to calculate thermodynamic quantities. Specifically, using Eq. (5) and the usual relation[12] between $U(t)$ and the partition function $Z$ we see that

$$Z(h) = \sum_n <n|\mathbf{R}_h(i\beta)U_{h=0}(i\beta)|n> \tag{6}$$

Here n labels states in Hilbert space. If $h(\tau) = h_\nu e^{i\nu\tau}$ with $\nu = 2\pi mT \neq 0$ and integer m then from Eq. (5) we see $R_h(i\beta) = 1$ and so we conclude from Eq. (6) that in this case $\partial^2 Z/\partial h_\nu^2 = 0$. This will serve as a consistency check on our subsequent results.

Note that at $\nu = 0$ (i.e. for static $h$), $R_h(i\beta) \neq 1$. However, in Appendix A we show by explicit calculation that for the model defined by Eq. (1), $\partial^2 Z/\partial h_0^2 = 0$. This differs from the result found for the nonlinear sigma model in two dimensions, where $\partial^2 Z/\partial h_0^2$ was shown to coincide with the physical susceptibility $lim_{q \to 0}[lim_{\omega \to 0}\chi(q,\omega)]$.

Equation 6 can also be used to derive the dependence on magnetic field of the effective action $S^{eff}$ describing the physics of low energy degrees of freedom if $S^{eff}$ is local in time. In this case $U$ can be represented as a Trotter product[13] and the quantity $R_h(\tau)R_h^{-1}(\tau)$ can be inserted at each time step yielding:

$$U_h(i\beta) = \prod_{j=1}^N R(i\tau_j)e^{H\Delta\tau}R^{-1}(i(\tau_j - \Delta\tau)) = \prod_{j=1}^N e^{H(h)\Delta\tau}) \tag{7}$$

For example, if the low energy theory is the nonlinear sigma model with action $(\partial_\tau n)^2 + (\nabla n)^2$, then it can be verified from Eq. (4) and Eq. (7) that this procedure leads to the replacement $(\partial_\tau n) \to \partial_\tau n + \vec{h} \times \vec{n}$ in the action, as it should.

We have not been able to extend the arguments summarized in the previous paragraph to the $z = 2$ critical theory defined in Eq. 1. The essential difficulty arises from the fact that we have integrated out gapless fermionic degrees of freedom to obtain $S_{eff}$. One consequence is that, as we shall see from the explicit calculations presented below, the coefficient coupling the field $h_{ext}(q,\Omega)$ to the fields $\phi$ is a nonanalytic function of $q$ and $\Omega$ as $q, \Omega \to 0$ and in particular depends strongly on the ratio $\Omega/q$. Such dependence is common in Fermi liquid theory and means that formulae such as Eq. (6), which are derived assuming spatially uniform fields with an arbitrary time dependence (i.e. $\Omega/q \to \infty$), may not yield the physical susceptibility, which corresponds to the opposite limit $q \to 0$ with $\Omega/q = 0$.

We turn now to an explicit calculation of $S^{eff}$ in the presence of an external field. Because the four spin-wave interaction is irrelevant, only two diagrams must be considered: these are generated by inserting one or two $h_{ext}$ vertices into Fig. 1a. The resulting diagrams are shown in Fig. 2. The analytic expression corresponding to these diagrams may be written:

$$S^{eff}(h_{ext}(q,i\Omega)) - S^{eff}(0) = iC_{k,\nu}(q,i\Omega)\epsilon_{abc}h_{ext}^a(q,i\Omega)\phi^b(Q+k,i\nu)\phi^c(-Q-k-q,-i\nu-i\Omega)$$
$$- \left[\frac{1}{2}D_{k\nu}^{(b)}(q,i\Omega) + (\delta^{ab} - \frac{1}{2})D^{(c)}(q,i\Omega)\right]h_{ext}^a(q,i\Omega)h_{ext}^a(-q-i\Omega) \tag{8}$$
$$\times \phi^b(Q+k,i\nu)\phi^b(-Q-k,-i\nu)$$

Here we have made the vector structure explicit and we have assumed that the fermion-fermion interaction producing the vertex corrections shown as the shaded areas in Fig. 2 is spin-independent. The coefficients C and $D^{(b,c)}$ are obtained by evaluating the diagrams shown in Fig. (2a) and Figs (2b,c) respectively. In the terms coming from Figs. (2b,c) we have retained only the contribution in which the two $h_{ext}$ fields and two $\phi$ fields have the same momentum and same direction, because this is the only term we shall need in our subsequent calculations.

¿From Eq. (8) we may evaluate $\chi$. Three diagrams arise; these are shown in Fig. 3. The resulting analytic expression is

$$\chi_u(q,\Omega) = 2T \sum_{k,\nu} C^2_{k\nu}(q,\Omega)\chi^2(k,\nu) - T\sum_{k,\nu}[3D^{(b)}_{k\nu}(q,\Omega) - D^{(c)}_{k\nu}(q,\Omega)]\chi(k,\nu) \qquad (9)$$

The sign of the term involving $C^2_{k,\nu}(q,\Omega)$ has two contributions: $i^2$ from the definition of the coupling constant and $-1$ from the cross product. In order to evaluate Eq. (9) we require expressions for the coefficients C and D. We will use general arguments based on the cancellation implied by Eq. (6) for $q=0$, $\Omega \neq 0$. We then use Fermi-liquid arguments to obtain the relation of this limit to the limit $\Omega = 0$, $q \neq 0$. We illustrate the general arguments with explicit calculations done in the weak coupling limit where all vertex corrections may be neglected.

The general expression for C is

$$C_{k\nu}(q,i\Omega) = 2T \sum_{p,i\epsilon} \Gamma(p+q,p)G(p,i\epsilon) \;\; [G(p+q,i\epsilon+i\Omega)\Lambda_{p,\epsilon}(k,\nu;q,\Omega) \qquad (10)$$
$$+ G(p+q,i\epsilon - i\Omega)\Lambda_{p,\epsilon}(k,-\nu;q,-\Omega)]$$

The vertex correction $\Gamma$ and block $\Lambda$ are defined in Fig. 2a. In the weak coupling limit, $\Gamma(p+q,p) = 1$ and

$$\Lambda_{p,\epsilon}(k,\nu;q,\Omega) = \frac{1}{i\epsilon + i\Omega + i\nu - \epsilon_{p+Q+k+q}} \qquad (11)$$

so

$$C_{k,\nu}(0,i\Omega) = 2[\chi^{-1}(k,i\nu + i\Omega) - \chi^{-1}(k,i\nu)]/i\Omega \qquad (12)$$

¿From Eq. (2) we find

$$C_{k,\nu}(0,i\Omega) = \frac{2i\,sgn(\nu)}{\Gamma} + i\nu C^{an}_{k,\nu} \qquad (13)$$

with $C^{an}(k,\nu)$ an analytic function of $\nu$ and $k^2$.

The usual Ward identity arguments[14] imply that Eq. (12) is in fact correct as $\Omega \to 0$ for any model of fermions with a short range spin-independent interaction. The point is that diagrams for C are generated by inserting a magnetic field vertex in diagrams for $\chi^{-1}$. At $q=0$ the magnetic field only shifts the energy of an electron line. The shift in energy may be absorbed into a shift in frequency and thus into a shift in external frequency, so a derivative with respect to magnetic field is equal to a derivative with respect to frequency.

A similar argument shows that $D^{(b)}$ and $D^{(c)}$ can be expressed in terms of the second frequency derivative of $\chi^{-1}$. Consider for example a diagram giving the leading singular contribution to $\chi^{-1}$ shown in Fig 4a. To obtain $D^{(b)}$ we assume that the external frequency is carried by the upper electron Green function and we differentiate twice with respect to the external frequency finding $D^{(b)} = 2\frac{\partial^2 \chi^{-1}}{\partial \nu^2}$. To obtain $D^{(c)}$ we assume that external frequency is equally shared between the upper and the lower lines. These considerations imply that $D^{(c)} = -2\frac{\partial^2 \chi^{-1}}{\partial \nu^2}$. The numerical factors in these equalities arise as follows: $D^b$ is defined to be the sum of the four diagrams of the type shown in Fig. 2b, while the second derivative of $\chi^{-1}$ produces two such diagrams if the external frequency is carried on the upper line; $D^{(c)}$ is the sum of two diagrams of the form shown in Fig. 2c, while the second derivative of $\chi^{-1}$ produces four diagrams of the type shown in Fig. 2b (if both derivatives act on the same line) and two of the type shown in Fig. 2c (if one derivative acts on an upper line and one on a lower line). The diagrams of the latter type enter with negative sign (because the external frequency flows backwards on the lower line) and there is a factor of $1/4$ multiplying all diagrams (because one half the external frequency flows on each line). Combining all factors we obtain:

$$3D^{(b)}_k(0,\Omega) - D^{(c)}_{k,\nu}(0,\Omega) = 8[\chi^{-1}(k,i\nu+i\Omega) - 2\chi^{-1}(k,i\nu) + \chi^{-1}(k,i\nu-i\Omega)]/(i\Omega)^2 \qquad (14)$$

or, using Eq. (2),

$$3D^b_{k,\nu}(0,\Omega) - D^{(c)}_{k,\nu}(0,\Omega) = 8\frac{|\nu+\Omega| - 2|\nu| + |\nu-\Omega|}{\Gamma(i\Omega)^2} + D^{an}(k,\nu) \qquad (15)$$

By substituting Eqs. (12) and (14) for C and D into Eq. (9) and using the discrete analogue of integration by parts, one sees immediately that $\chi(0,\Omega) \equiv 0$, consistent with Eq. (6).

Because it will be important in subsequent arguments, we now discuss the origin of the nonanalytic frequency dependence in more detail. Consider first the term $|\nu|/\Gamma$ in $\chi^{-1}$. The usual Landau threshold arguments[15] imply that the nonanalytic behavior of $\chi^{-1}$ comes from diagrams such as that shown in Fig. (4a), which may be divided into two disconnected parts by cutting two electron lines. The usual results of Fermi liquid theory imply that the vertex $\Gamma_Q$ coupling an antiferromagnetic spin fluctuation to the fermions and defined in Fig. 4 is an analytic function of $k, \Omega$. The nonanalytic term $|\nu|/\Gamma$ comes from the product $G(p, \epsilon) G(p+Q+k, \epsilon+\nu)$ and in particular comes from the region of the $p, \epsilon$ integrals where both $p$ and $p+Q$ are near the Fermi surface and $\epsilon$ is small. If more lines were present to share the frequency $\nu$, the nonanalyticity would be washed out. The general Ward identity says that in the $\Omega$-limit inserting a field vertex is equivalent to differentiating with respect to frequency. Therefore, the nonanalytic contributions to C and D come from putting $h_{ext}$ vertices on the isolated electron lines in Fig. (4a), as shown in Figs. (4b-d).

We now turn to the limit $q \neq 0$, $i\Omega = 0$ which corresponds to the physically measured bulk susceptibility. Instead of calculating $C(q,0)$ directly we combine our previous result for $C(0,\Omega)$ with a calculation of

$$\Delta C = C(q,0) - C(0,\Omega) \qquad (16)$$

Because the entire nonanalytic contribution to $C(0,\Omega)$ is given by diagrams of the form shown in Fig. (4b), it is convenient to write

$$C(q,\Omega) = C_{4b}(q,\Omega) + C_{rest}(q,\Omega) \qquad (17)$$

with $C_{4b}$ given by the sum of all diagrams of the form shown in Fig. (4b) and $C_{rest}$ given by the sum of all other diagrams. We now calculate separately the contributions $\Delta C_{4b}$ and $\Delta C_{rest}$. We begin with $\Delta C_{4b}$, which is

$$\Delta C_{4b} = T \sum_{p,\epsilon} [G(p+q,\epsilon)G(p,\epsilon)\Gamma(p+q,p) - G(p,\epsilon+\Omega)G(p,\epsilon)\Gamma(p+q,p)] \qquad (18)$$

$$\cdot \Gamma_Q^2 G(p+Q+k, \epsilon+\nu)$$

Because the factor in brackets is convergent we may perform the p-integral first. In this integral the factor $G(p+q,\epsilon)G(p,\epsilon)$ appearing in the first term contributes only for momenta $p$ far from the Fermi surface; for such momenta the $\nu, k$ dependence is smooth. Therefore the first term in brackets gives only an analytic contribution to $\Delta C_{4b}$. The second term in brackets is precisely $C_{4b}(0,\Omega)$. Therefore the nonanalyticity of $\Delta C_{4b}$ precisely cancels the nonanalyticity in $C_{4b}(0,\Omega)$ so that $C_{4b}(q,0)$ is analytic. $C_{rest}(q,0)$ is also analytic, for the same reason that $C_{rest}(0,\Omega)$ is analytic. This completes the argument that $C(q,0)$ is an analytic function of $\nu$. Because it must be odd under time reversal we have

$$\Delta C^{an} = iC_1 \nu + ... \qquad (19)$$

Here $C_1$ is a non-universal constant set by short length scale physics and the ellipsis denotes terms of order $\nu k^2$ and $\nu^3$.

Very similar arguments apply to D. The leading singular nonanalytic contribution to $D(0,\Omega)$ is cancelled by terms in $\Delta D$, so $D(q,0) \sim$ constant. There is one subtlety: among the diagrams in $D_{rest}$ are those with one field insertion on an isolated G-line in $\chi^{-1}$ and one in $\Gamma_Q$. In the $\Omega$-limit, the insertion in $\Gamma_Q$ gives a contribution $d\Gamma_Q/d\nu$. But because $\Gamma_Q$ is an even analytic function of $\nu$, this gives a factor $\nu$, and therefore a contribution of order $|\nu|$ to D. This is however negligible relative to the leading term, which is a constant. We therefore write $3D^{(b)}(q,0) - D^{(c)}(q,0) = D$. Substituting into Eq. (9) one finds

$$\lim_{q \to 0} \chi(q,0) = 2T \sum_{k,\nu} [iC_1 \nu]^2 \chi^2(k,\nu) - T \sum_{k,\nu} D \chi(k,\nu) \qquad (20)$$

Using Eq. (1) in Eq. (20) we find

$$\chi(q \to 0, 0) = -2C_1^2 T \sum_{k,\nu} \frac{\nu^2}{\left(\frac{|\nu|}{\Gamma} + k^2 + \delta\right)^2} - T \sum_{k,\nu} \frac{D}{\frac{|\nu|}{\Gamma} + k^2 + \delta} + ... \qquad (21)$$

Eq. (21) may be evaluated in different limits. One particularly important limit is the "quantum critical regime" in which $\delta \lesssim T$ so that the critical behavior is cut off only by thermal fluctuations. It has been claimed[4,8] that the underdoped high $T_c$ superconductors are in this regime. Evaluating Eq. (21) in two dimensions, assuming $\delta \lesssim T$ yields, to logarithmic accuracy,

$$\chi(q \to 0, 0) = \chi_{bg} + C_2 \delta - \frac{D}{4\pi} T \ln[T/\delta] \qquad (22)$$

Here the positive constant $C_2$ is determined by high energies where even Eq. (2) is not valid.

In the quantum critical regime of the $d = 2$, $z = 2$ theory, it has been shown[5] that $\delta \sim T/\ln[T]$ so that the dependence of $\chi$ on $T$ is, for all practical purposes, linear.

In the $d = 3$ quantum critical regime the $\delta$ dependence is not important because the $k$ integral is convergent even at $\nu = 0$. Evaluating Eq. (21) by scaling one finds $\chi(q \to 0, 0) = \chi_0 + \alpha D T^{3/2} + ...$ where $\alpha$ is a constant not determined by the scaling argument.

We now discuss the sign and magnitude of $D(q, 0)$. We have calculated $D$ by evaluating Figs. (2b) and (2c) for noninteracting fermions with spectrum $\epsilon_p$ and spin fluctuation vertex $\Gamma_Q$. We obtain, after performing the frequency integration,

$$\tilde{D}^{(b)}(Q, 0) = 8 \int \frac{d^2 p}{(2\pi)^2} \frac{f(\epsilon_p) - f(\epsilon_{p+Q}) + (\epsilon_{p+Q} - \epsilon) f'(\epsilon_p) + \frac{1}{2}(\epsilon_{q+Q} - \epsilon_p)^2 f''(\epsilon_p)}{(\epsilon_p - \epsilon_{p+Q})^3} \qquad (23)$$

$$\tilde{D}^{(c)}(Q, 0) = -4 \int \frac{d^2 p}{(2\pi)^2} \frac{2(f(\epsilon_p) - f(\epsilon_{p+Q})) + (\epsilon_{p+Q} - \epsilon_p)(f'(\epsilon_p) + f'(\epsilon_{p+Q}))}{(\epsilon_p - \epsilon_{p+Q})^3} \qquad (24)$$

Here $\tilde{D}^{(b,c)} = D^{(b,c)}/\Gamma_Q^2$. We have evaluated these integrals for tight binding bandstructures with dispersion

$$\epsilon_p = -2t(\cos p_x + \cos p_y) - 4t' \cos p_x \cos p_y \qquad (25)$$

and various ratios of $t'/t$. Some representative results are shown in Fig 5. We find that $D = 3D^{(b)}(Q, 0) - D^{(c)}(Q, 0)$ is positive for $|Q| < 2p_F(\theta)$, negative for $|Q| > 2p_F(\theta)$. For $|Q|$ differing by more than $\approx 20\%$ from the critical value $2p_F(\theta)$ we find $D/(4\pi)$ is of order $0.03\Gamma_Q^2/t^3$. In a Fermi liquid one expects $\Gamma_Q$ to be set by an electron-electron interaction energy which is presumably $\sim t$, so $D/(4\pi) \lesssim 0.03/t$.

The sign of $D$ can be understood from the following qualitative arguments. First, from Eqs. (23,24) we see that if $Q > 2p_F$, $D < 0$ because the terms involving derivatives of the Fermi function vanish and the sign of the denominator is opposite to that of the numerator. Of course, $Q > 2p_F$ corresponds to undamped spin waves, which contradicts our starting assumption. Suppose now $Q < 2p_F$. Because a Fermi liquid has no scale other than $2p_F$, the box diagrams must be smooth functions of the external momenta if these momenta are less than $2p_F$. Therefore it is reasonable to suppose that the sign of the diagram in 2b (when two external lines are at $q \approx 0$) is the same as 1b (when all four external lines are at $Q$). But the sign of the diagram in 1b is positive, corresponding to repulsive spin-fluctuation interactions and the stability of the theory defined by Eq. (1). Of course this argument is merely qualitative. It may be possible to devise band structures for which $D < 0$ for $Q < 2p_F$.

To summarize, we have shown that in a two-dimensional Fermi liquid which undergoes a zero temperature $z = 2$ antiferromagnetic transition the uniform susceptibility in the quantum critical regime is constant plus a term which is approximately linear in $T$. The coefficient and the sign of the leading $T$-dependence is nonuniversal. This leaves open the possibility that in real materials the temperature dependence of $\chi(q \to 0, 0)$ is determined by the terms which are non-leading in the $z = 2$ theory; for example the terms proportional to $C_2$ in Eq. (22) which are small by a factor of $\delta/T$. A discussion of these terms requires physics at high energy scales which is beyond the scope of the $z = 2$ critical theory studied here.

However, for band structures similar to those proposed for the high $T_c$ superconductors and for wavevectors $Q$ not too close to $2p_F$, the sign is negative and the magnitude is small. The behavior we find differs in several respects for the behavior of $\chi(q \to 0, 0)$ near the $z = 1$ transition discussed previously[8]. In the $z = 1$ case the transition is between a magnetically ordered phase with linearly dispersing undamped spin waves and a singlet phase with a gap to all spin excitations. The background term $\chi_{bg}$ vanishes at $T = 0$ at the critical point and in the singlet phase, and the leading term in the quantum critical regime is $\chi(q \to 0, 0) \sim AT$ with A a universal and positive number. Note that if the ordering wavevector were outside the particle-hole continuum ($Q > 2p_F$) the spin fluctuation contribution to $\chi(q \to 0, 0)$ would be given by the universal $z = 1$ results but there would in addition be a fermion contribution described by Eq. (22) but with $D < 0$. The resulting $\chi(q \to 0, 0)$ might be in qualitative agreement with high $T_c$ data.

To conclude, we comment on the relationship of our result to previous work of Sachdev and to data. Sachdev has combined the rotation arguments discussed above Eq. (6) with finite size scaling to argue that near a quantum critical point the free energy in an SU(2) invariant system is a universal function $f_u(h_{ext}/T)$ and that the "Wilson ratio" (susceptibility divided by specific heat coefficient) takes a universal value[10]. Our results do not contradict those of Sachdev; rather, they imply that in the physical limit $\Omega/q \to 0$, $f_u(h_{ext}/T) \equiv 0$, a possibility allowed by the general arguments. The mathematical reason for the vanishing of $f_u$ is the nonanalyticity in $q, \Omega$ of the coefficients coupling $h_{ext}(q, \Omega)$ to the ordering fields $\phi$. In the limit $\Omega/q \to \infty$ where the rotation arguments apply the couplings have the expected form. However, in the limit $\Omega/q \to 0$ which is relevant to measured bulk susceptibilities, the leading (universal) terms in the couplings vanish. The physical reason underlying this mathematics was stated previously[5]: the universal results are ultimately derived from the fact that a uniform magnetic field generates rotations of, e.g. the staggered magnetization. However, in $z = 2$ problems the staggered magnetization decays into a particle-hole pair before it has time to precess significantly, and so arguments based on precession do not give the leading behavior of $\chi(q \to 0, 0)$. A similar vanishing of the universal function is found in a class of impurity models[16].

Our results also justify the action Eq. (1) used in previous work[1,5] to analyze $z = 2$ transitions. This action neglected e.g. triangle graphs such as Fig. (1b) in which one vertex is the nearly uniform magnetization and two are the staggered magnetization. Our analysis implies that if these terms are included in $S^{eff}$ (eq. (1)) and then the uniform magnetization fluctuations are integrated out, the resulting contributions are nonleading.

Concerning experiments, we note that the observed linear $T$-dependence of the uniform susceptibility in high-$T_c$ superconductors has been argued to be evidence of $z = 1$ critical behavior. We see that would be rather difficult to explain in a $z = 2$ model, because for reasonable band structures the coefficient of the leading $T$-dependent term is negative and of small magnitude if the ordering wavevector is inside the particle-hole continuum; however, there are non-leading (by a factor $\delta\Gamma/T$) terms of positive sign which could explain the data.

*Note added:* After this research was completed we received a preprint from S. Sachdev, A. Chubukov and A. Sokol reporting the study of the crossover from $z = 1$ to $z = 2$ behavior[17]. They obtain somewhat similar results for $\chi(q \to 0, 0)$ but they do not consider logarithms and so do not distinguish between leading and non-leading terms.

## ACKNOWLEDGEMENTS

We thank A. W. W. Ludwig for helpful conversations and S. Sachdev and A. Chubukov for useful remarks and questions and especially for pointing out to us an error in our analysis of the non-leading $C$-vertex. Part of the work reported here was performed at the Aspen Center for Physics.

## APPENDIX: APPENDIX

In this Appendix we evaluate Eq. (6) at $T > 0$ for a static uniform magnetic field $\vec{h}_{ext} = h\hat{z}$. Equation (6) may be written as a coherent state path integral,

$$Z(h) = \int \mathcal{D}\vec{\phi}^* \mathcal{D}\phi \exp -T \sum_k \left[ \int_0^\beta d\tau \int_0^\beta d\tau' K(\tau - \tau') \vec{\phi}_k(\tau) \cdot \vec{\phi}^*_{-k}(\tau') \right.$$

$$\left. + \int_0^\beta d\tau (k^2 + \delta) \vec{\phi}_k(\tau) \cdot \vec{\phi}^*_k(\tau) \right] \tag{A1}$$

The fields $\phi$ obey the rotated boundary conditions

$$\phi^a(\beta) = R^{ab}(\beta)\phi^b(0) \tag{A2}$$

The kernel $K(\tau)$ is given by

$$K(\tau) = T \sum_{\nu < \Gamma} e^{i\nu\tau} |\nu|/\Gamma \tag{A3}$$

Because we have chosen the field to be in the $z$ direction, the functional integral over $\phi^z, \phi^{*z}$ is $h$-independent and so may be ignored. To perform the integral over the other components we define new fields $\psi^x, \psi^y$ by

$$\psi^x_k(\tau) = \cosh(h\tau)\phi^x_k(\tau) - i\sinh(h\tau)\phi^y_k(\tau) \tag{A4}$$

$$\psi^y_k t(\tau) = i\sinh(h\tau)\phi^x_k(\tau) + \cosh(h\tau)\phi^y_k(\tau) \tag{A5}$$

We similarly define $\psi^{*x}$ and $\psi^{*y}$ using the complex conjugate of Eqs. (A4,A5). The transformation (A4,A5) has unit determinant, so $\mathcal{D}\phi\mathcal{D}\phi^* = \mathcal{D}\psi\mathcal{D}\psi^*$. ¿From Eqs. (A2, A4,A5) and Eq. (5) we see $\psi(\beta) = \psi(0)$, so the $\psi$ fields may be expanded in Matsubara frequencies in the usual way, $\psi(\tau) = T\sum_\nu e^{i\nu\tau}$.

We now substitute Eq. (A4,A5) into Eq. (A1), expand the fields $\psi$ in Matsubara frequencies, use Eq. (A3) and perform the $\tau$ integrals. The only $h$-dependence is in the first (double integral) term of the argument of the exponential in Eq. (A1). Denoting this term by $\sum_k S_k$ we find

$$S_k(h) = T^2 \sum_{\nu_1\nu_2\nu_3} \frac{|\nu_3|}{\Gamma}(2 - 2\cosh\beta h)[(\psi^x(\nu_1)\psi^{*x}(\nu_2) + \psi^y(\nu_1)\psi^{*y}(\nu_2))f(\nu_1,\nu_2,\nu_3;h)$$

$$+(\psi^x(\nu_1)\psi^{*y}(\nu_2) + \psi^{*x}(\nu_1)\psi^y(\nu_2)g(\nu_1,\nu_2,\nu_3;h)] \tag{A6}$$

with

$$f(\nu_1,\nu_2,\nu_3;h) = +\frac{2[(\nu_1+\nu_3)(\nu_2+\nu_3) - h^2]}{((\nu_1+\nu_3)^2 + h^2)((\nu_2+\nu_3)^2 + h^2)} \tag{A7}$$

$$g(\nu_1,\nu_2,\nu_3;h) = -\frac{h[\nu_1+\nu_2+2\nu_3]}{((\nu_1+\nu_3)^2 + h^2)((\nu_2+\nu_3)^2 + h^2)} \tag{A8}$$

The first term (involving $F$) can contribute to $Z$ to order $h^2$ only if $\nu_1 = \nu_2$. It is convenient to distinguish the two cases $-\nu_3 = \nu_1 = \nu_2$ and $-\nu_3 \neq \nu_1 = \nu_2$. One finds:

$$S_k^1(h) = \left(1 - \frac{(\beta h)^2}{12}\right)\sum_\nu \frac{|\nu|}{\Gamma}(\psi_k^x(\nu)\psi_k^{*x}(\nu) + \psi_k^y(\nu)\psi_k^{*y}(\nu))$$

$$-h^2 \sum_\nu \sum_{\nu_3 \neq -\nu} \frac{|\nu_3|}{\Gamma}\frac{1}{(\nu+\nu_3)^2}(\psi_k^x(\nu)\psi_k^{*x}(\nu) + \psi_k^y(\nu)\psi_k^{*y}(\nu)) \tag{A9}$$

The second term (involving $g$) plainly vanishes if $\nu_1 = \nu_2 = -\nu_3$ and is $O(h^3)$ if $\nu_1 \neq \nu_2 \neq -\nu_3$. We may therefore write

$$S_k^2(h) = \frac{h}{\Gamma}\sum_{\nu_1 \neq \nu_2} \frac{|\nu_1| - |\nu_2|}{(\nu_1-\nu_2)}(\psi_k^x(\nu_1)\psi_k^{*y}(\nu_2) + \psi_k^{*x}(\nu_1)\psi_k(\nu_2)) \tag{A10}$$

¿From Eqs. (A1,A9,A10) we obtain

$$Z(h) - Z(h=0) = (\beta h)^2 \cdot \left[\frac{1}{6}\sum_{\nu,k}\frac{|\nu|/\Gamma}{\frac{|\nu|}{\Gamma} + k^2 + \delta}\right.$$

$$+2T^2 \sum_\nu \sum_{\nu_3 \neq \nu} \frac{|\nu_3|}{\Gamma}\frac{1}{(\nu-\nu_3)^2}\frac{1}{\frac{|\nu|}{\Gamma} + k^2 + \delta} \tag{A11}$$

$$-\frac{2}{\Gamma^2}T^2 \sum_{\nu_1}\sum_{\nu_2 \neq \nu_1}\left[\frac{|\nu_1|-|\nu_2|}{(\nu_1-\nu_2)}\right]^2 \frac{1}{\frac{|\nu_1|}{\Gamma}+k^2+\delta}\frac{1}{\frac{|\nu_2|}{\Gamma}+k^2+\delta}$$

By use of the identity

$$\frac{1}{\frac{|\nu_1|}{\Gamma}+k^2+\delta}\frac{1}{\frac{|\nu_2|}{\Gamma}+k^2+\delta} = \frac{\Gamma}{|\nu_1|-|\nu_2|}\left(\frac{1}{\frac{|\nu_2|}{\Gamma}+k^2+\delta} - \frac{1}{\frac{|\nu_1|}{\Gamma}+k^2+\delta}\right) \tag{A12}$$

along with the results $\nu_1 = 2\pi nT$ and

$$\sum_{n=1}^\infty \frac{1}{n^2} = \frac{\pi^2}{6} \tag{A13}$$

one may show that the last term precisely cancels the first two, so $Z(h) - Z(h=0) = 0$.

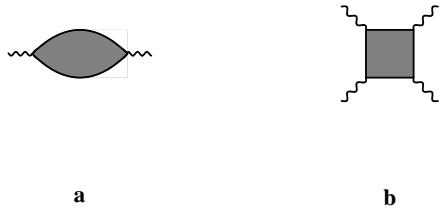

a      b

Fig 1

FIG. 1. Diagrams contributing to the effective action for spin fluctuations, Eq. 1. The wavy line denotes a spin fluctuation and the shaded regions denote fermion two and four point functions which are irreducible with respect to the fermion-fermion interaction.

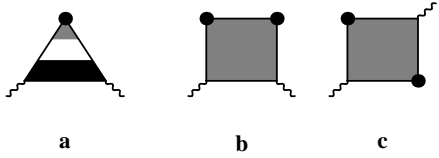

Fig 2

FIG. 2. Diagrams describing the coupling between a nearly uniform external magnetic field (heavy dot) and fluctuations of the staggered magnetization (wavy line). The solid line denotes the electron propagator $G$ and the shaded regions denote dressing of the diagram by the fermion-fermion interaction. Each diagram is understood to be the sum of two diagrams with opposite direction of electron lines. In Fig. 2a we have picked out the two fermion-magnetic field vertex $\Gamma(p+q,p)$ (lightly shaded triangle) and the two fermion-two spin fluctuation vertex (heavily shaded trapezoid), because these quantities are needed for the discussion in the text. We have not similarly decomposed Figs. 2b and 2c because detailed expressions for these diagrams are not needed.

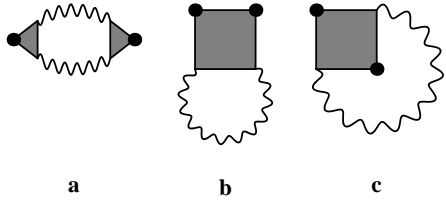

Fig 3

FIG. 3. Diagrams used in computation of contribution of spin fluctuations (wavy line) to uniform susceptibility. The shaded triangle in Fig. 3a is the vertex shown in Fig. 2a; the squares in 3b,c are the vertices shown in Figs. 2b,c.

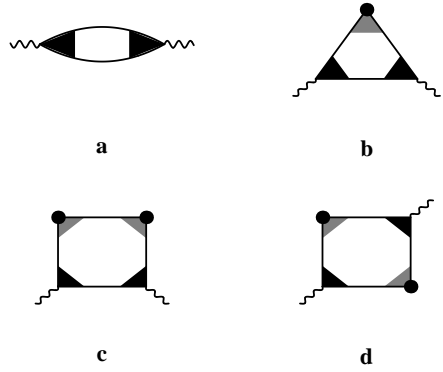

**a**

**b**

**c**

**d**

Fig 4

FIG. 4. Diagrams contributing to nonanalytic terms in $\chi, C$, and $D$. The heavy dot denotes the nearly uniform external field, the wavy line denotes the spin fluctuations, and the solid lines the dressed electron propagators. The shaded triangle is the vertex $\Gamma(p+q,p)$ coupling the dressed electron propagators to the external magnetic field and the solid triangle is the vertex $\Gamma_Q$ coupling the dressed electron propagators to the antiferromagnetic spin fluctuations.

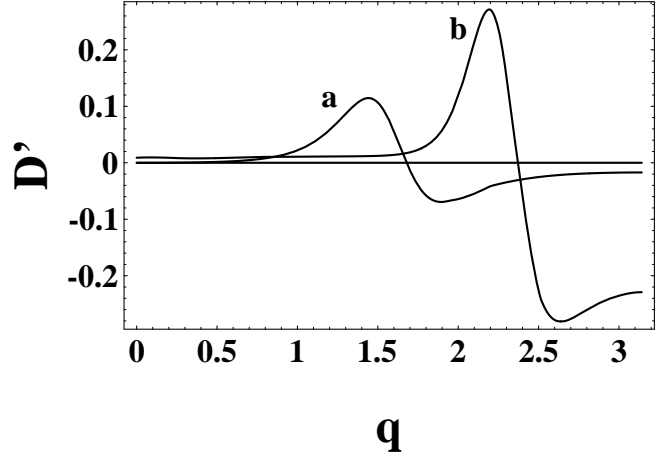

Fig 5

FIG. 5. Plot of coefficient $D' = D(\pi\hat{x} + q\hat{y}, 0)/(4\pi\Gamma_Q^2)$ calculated for noninteracting electrons with the dispersion given in Eq. (25), $t = 1$, $t'/t = -0.1$ and two electron densities: $n = 0.5$ (curve (a)) and $n = 0.7$ (curve (b)).